\documentclass{IEEEtran}
\usepackage{graphics}
\usepackage{epsfig}
\usepackage{tikz}
\newcommand{\ba}{\begin{array}}
\newcommand{\ea}{\end{array}}

\renewcommand{\a}{\alpha}
\newcommand{\Si}{\Sigma}

\renewcommand{\l}{\lambda}
\newcommand{\bbibl}{\begin{thebibliography}{99}}
\newcommand{\ebibl}{\end{thebibliography}}
\usetikzlibrary{shapes}
\usepackage{verbatim}
\newcommand{\Ber}{\mbox{Ber}}

\newcommand{\Bin}{\mbox{Bin}}
\newcommand{\Normal}{\mbox{Normal}}
\newcommand{\Event}{\mbox{Event}}

\title{\bf {Hypothesis Testing and Decision Theoretic Approach for Fault Detection in Wireless Sensor Networks}}

\author{
Mrinal Nandi,$^1$  Amiya Nayak, $^2$ Bimal Roy,$^3$ Santanu Sarkar $^3$
\\[10pt]
$^1$  Department of Statistics, West Bengal State University, Barasat, West Bengal, India.\\
$^2$  SITE, University of Ottawa, 800 King Edward Ave., Ottawa, ON K1N 6N5, Canada.\\
{e-mail : mrinal.nandi1@gmail.com (corresponding author$^1$ ). }   \\
$^3$  ASD, Indian Statistical Institute, 203 B. T. Road, Kolkata-700 108, India. \\
}

\begin{document}
\maketitle

\begin{abstract}
Sensor networks aim at monitoring their surroundings for event detection and object tracking. But due to failure or death of sensors, false
signal can be transmitted. In this paper, we consider the problem of fault detection in wireless sensor network (WSN), in particular, 
addressing both the noise-related measurement error and sensor fault simultaneously in fault detection. We assume that the sensors are
placed at the center of a square (or hexagonal) cell in region of interest (ROI) and, if the event occurs, it occurs at a particular cell 
of the ROI. We propose fault detection schemes that take into account error probabilities into the optimal event detection process. 
We develop the schemes under the consideration of Neyman-Pearson test and Bayes test.
\end{abstract}


\section{Introduction}
Traditional and existing {\it sensor-actuator networks} use wired communication, whereas wireless sensor networks provide radically new communication and networking paradigms and myriad new applications. The wireless sensors have small size, low battery capacity, non-renewable power supply, small processing power, limited buffer capacity and low-power radio. They may measure distance, direction, speed, humidity, wind speed, soil makeup, temperature, chemicals, light, and various other parameters.

Recent advancements in wireless communications and electronics have enabled the development of low-cost WSN. A WSN usually consists of a large number of small sensor nodes, which are equipped with one or more sensors, some processing circuit, and a wireless transceiver. One of the unique features of a WSN is random deployment in inaccessible terrains and cooperative effort that offers unprecedented opportunities for a broad spectrum of civilian and military applications; such as industrial automation, military tactical surveillance, national security, and emergency health care~\cite{ASSC,PK,A}. Sensor Networks are useful in detecting topological events such as forest fires ~\cite{FSAW}.

Sensor networks aim at monitoring their surroundings for event detection and object tracking~\cite{ASSC,MS}. Because of this surveillance goal, {\it coverage} is the functional basis of any sensor network. In order to fulfill its designated surveillance tasks, a sensor network must fully cover the Region of Interest (ROI) without leaving any {\it internal sensing hole}~\cite{B,BS,C,FKP}. However, it cannot be expected that sensors are placed in a desired way at initiation as they are often randomly dropped due to operational factors. Furthermore, a sensor could die or fail at runtime for various reasons such as power depletion, hardware defects etc. In wireless sensor and actuator networks (WSAN), sensors can be placed by mobile actuator(s), i.e., robot(s). If sensors are mobile, they can place themselves without any external help. But as physical movement consumes a large amount of energy for the sensor nodes, a movement assisted sensor placement scheme is preferred~\cite{BHE, FLN,LNS}. On the other hand, some sensors could be marked redundant in terms of local sensing coverage and that sensors are called {\it passive sensors}. Passive sensors could be either deployed on purpose or determined by area coverage protocol~\cite{G}.

So far, a number of movement-assisted sensor placement algorithms have been proposed. An exclusive survey on these topics is presented by Li et al.~\cite{L}. On the other hand sensor could die or fail at runtime for various reasons such as power depletion, hardware defects etc. So, even after the ROI is fully covered by the sensors, wrong information can send by some sensors or sensors may fail to detect the event due to noise or obstructions. Chen et al.~\cite{CKS} proposed a localized fault detection algorithm for WSN. Sharma et al.~\cite{S} characterize different types of faults and fault detection methods.

Mousavi et al.~\cite{M} presented a distributed one step deployment (OSD) algorithm. This algorithm partitions the ROI evenly into two-dimensional square grids, and instructs sensors to occupy all the grid points. The intuition is that if every grid point is occupied by a sensor, then the ROI is fully covered, and the sensors form a connected network. Fletcher et al.~\cite{F} present randomized algorithms using more than one robot for coverage repair in WSN. They propose two algorithms for grid based ROI and simulate the path traveled by the robots for different values of parameters (number of sensors, number of robots etc.).

One of the important sensor network applications is monitoring inaccessible environments. Sensor networks are used to determine event regions and boundaries in the environment with a distinguishable characteristic~\cite{KI,CG,NM}. The basic idea of distributed detection~\cite{T} is to have each of the independent sensors make a local decision (typically, a binary one, i.e., whether or not an event has occurred) and then combine these decisions at a \emph{fusion sensor} (a sensor which collects the local information and takes the decision) to generate a global decision or send these information to the base station. Optimal distributed design have been sought under both the Bayesian and the Neyman-Pearson performance criteria~\cite{V}.

\subsection{Our Motivation}
In this paper, we are interested in the following: determining an event and the position of the event in a environment where that event may have occurred. We assume that ROI is partitioned into suitable number of squares (i.e., we consider ROI as a rectangular grid with square cells. We also consider regular hexagonal grid with regular hexagonal cells in a separate section). We also assume that sensors have already been placed at the centers (which we call them \emph{nodes}) of the square cells.

One fundamental challenge in the event detection problem for a sensor network is the detection accuracy which is limited by the amount of noise associated with the measurement and the reliability of sensor nodes. The sensors are usually low-end inexpensive devices and sometimes exhibit unreliable behavior. For example, a faulty sensor node may issue an alarm even though it has not received any signal for event, or it cannot detect an actual event and vice versa. Moreover, a sensor may be dead in which case sensor cannot send any alarm.

The event region can be large, and if an event occurs at a particular point of the region then the sensor cannot determine exactly where the event has happened. There are cases where a fusion sensor cannot make a decision. Consider, for example, a network of sensors that are capable of sensing mines or bombs. If we assume that either no mines (or bombs) are placed or very few mines (or bombs) are placed on a particular area of ROI, then an important query in this situation could be whether bombs are placed or not. In that case, fusion sensor does not help to take the decision. All sensors have to communicate with the base station, and base station will take the decision about the query.

\subsection{Assumptions}
In this section, we describe the assumption we make; some are new and some are identical to the ones made by other researchers. The new assumptions lead to a new type of problem statement and a new approach to solve the detection problem.

\begin{itemize}

\item We assume that if an event occurs in the square where the particular sensor (call it as center node) lies then that particular sensor can detect the event with a high probability whereas if an event occurs in the adjacent square of the center square then the sensor can detect the event with a lesser probability. Moreover, the probability of detection decreases as distance between the sensor and event square increases. Hence, only one node (center node of the event square) can detect the event square with the highest probability, say $p_1$, four distance-one nodes can detect the event square with lower probability, say $p_2$, and the four distance-two nodes can detect the event square with lowest probability, say $p_3$. Here, distance-one node means the node which is placed at the center of an adjacent square (i.e., a side is common with the center square) and distance-two node means that the node which placed at the center of an square which has a common vertex with the center square. We assume that no other sensor can detect the event square, and $p_1 > p_2 > p_3$.

\item Unlike the previous work, we assume that if the event occurs then it occurs at only one particular square of the grid which will be known as event square, and there is no fusion sensor.

\item Sensors are deployed or manually placed over ROI in such a way that they cover the entire ROI. We assume that sensors are placed priori at the center (which are known as nodes) of every square cell (we also consider regular hexagonal grid with regular hexagonal cell in a separate section).

\item Each sensor node can determine its location through beacon positioning mechanism~\cite{BHE}. Sensors are also able to communicate with the base station. Unlike the previous work, we assume that either the event occur at one particular square of the grid which will be known as {\it event square} or event does not occur (in that case we say ROI is {\it normal}).

\item We assume that there is no fusion sensor (the sensor which will take decision locally); all sensors  communicate with the base station which takes the decision.

\item Sensor at the center of the event square can detect the event square with highest probability and the sensors which are situated at a certain distance from the event square can detect the event square with lesser probability (due to different noise, distance, obstructive, etc.). We also assume that there is prior probability of a particular square to be an event square. Even if event square is detected by a sensor, it may not respond or send the information to the base station due to some technical fault (we call that sensor as {\it faulty sensor}) with some probability. Conversely, if an event square was not detected or if there is no event square (i.e., normal situation) then also a faulty sensor can falsely respond or send the wrong information to the base station with some probability. A sensor is called a {\it dead sensor} if the sensor does not work at all. A dead sensor sends no response in either case. If a sensor is dead or the ROI is normal then the sensors send no information, i.e., do not respond. We also assume that the sensors work independently, i.e., detection and response of different sensors are independent.

\end{itemize}

\subsection{Our Contribution}
In our theoretical analysis, we propose fault detection schemes that take into account error probabilities in the optimal event detection process. We develop the schemes under the consideration of classical hypothesis testing and Bayes test. We calculate different error probabilities and find some theoretical results involving different parameters such as probability of false alarm of a sensor, probability of event detection by a sensor, prior probability of occurring a event, etc. Finally, we calculate different error probabilities, Bayes test and Neyman-Pearson Most Powerful(MP) test for some specific values of the parameters and state some concluding remarks analyzing the calculation results.

Parameters $p_1, p_2, p_3$, the detection probabilities of a sensor, and the probability of sending information correctly by a sensor, cannot be estimated from the real life situation but we can estimate them experimentally beforehand. The prior probability of the event cannot be estimated. In various situations, it may be known in which case we apply Bayes test; otherwise, we use Neyman-Pearson MP test.

In this paper, we propose a rule for the base station to take a decision compiling the information coming from the all sensors and find the optimal solutions. We consider two type of error: type I error when an event occurs but the sensor report normal (which is the more serious error) and type II error when ROI is normal but sensor report as an event. We observed that type I and type II errors decrease when detection probabilities increase. If detection probabilities are low then type I error is close to $1$.  If probability of occurrence of the event is high but detection probabilities are small then type I and type II errors are high, which means there is no use of sensors. So, when the probability of occurrence of the event is high, we have to use sensors with high detection probability (i.e. sensors with much better quality). We calculate the MP test and the Bayes test for some specific values of the parameters. We observed that for small values of detection probability and large value of loss, the Bayes test is not applicable. When loss is large, we cannot use sensors with small detection probabilities to decide about the event square using Bayes test. We also observed that when the size of the test is small we cannot use sensors with small detection probabilities for MP test; we have to use good sensors (sensor with high detection probability) for MP test in this case.

In a separate section, we have considered ROI as a hexagonal grid, i.e., it partitioned into suitable number of congruent regular hexagons and sensors are placed at the center of the hexagons. Instead of three detection probabilities (as in the case of square grid), we assume there are two detection probabilities $p_1, p_2$ with $p_1>p_2$, same for all hexagons. Other assumptions are same.

\section{Related Work}
There are several papers which consider only the coverage problem of sensor networks. Tseng and Huang~\cite{TH} formulate the problem as a decision problem, whose goal is to determine whether every point in the service area of the sensor network is covered by at least $k$ sensors, where $k$ is a predefined value.

Lou et al.~\cite{LDH} consider two important problems for distributed fault detection in WSN: 1) how to address both the noise-related measurement error and sensor fault simultaneously in fault detection and 2) how to choose a proper neighbourhood size $n$ for a sensor node in fault correction such that the energy could be conserved. They propose a fault detection scheme that explicitly introduces the sensor fault probability into the optimal event detection process. They show that the optimal detection error decreases exponentially with the increase of the neighbourhood size.

Filippou et al.~\cite{FKP} measure the ability of the network to interact with observed phenomena taking place in the ROI. In addition, coverage is associated with connectivity and energy consumption, both of which are important aspects of the design process of a WSN. The paper aims at offering a critical overview and presentation of the problem as well as the main strategies developed so far.

An example of a uniform deterministic coverage is a grid based sensor deployment where sensors are located on the intersection points of a grid. This requires manual placement, which is realistic for small number of nodes, and an accessible environment. This placement ensures complete coverage of the field with the minimum number of sensors. The minimum number of sensors needed to cover an area is given by Williams~\cite{W}.

Nandi and Li~\cite{MX} consider coverage problem in wireless sensors and actuator network composed of static sensors dropped stochastically in the ROI which is a rectangular grid with square cells. Sensors are dropped at the vertices of the grid from air. Sensors have communication radius $r_c$ and sensing radius $r_s$, where $r_c \geq \sqrt{2} r_s$. Actuator can take, carry and place the sensors according to some pre-assigned algorithm. Nandi and Li \cite{MX} developed three algorithms for the actuator under two different conditions and compared these algorithms in context with some pre-assigned parameters. They also deduced some theoretical results on the parameters and some simulation results applying the three algorithms.

Krishnamachari and Iyengar~\cite{KI} propose a distributed solution for canonical task in WSN, i.e., the binary detection of interesting environmental events. They explicitly take into account the possibility of sensor measurement faults and develop a distributed Bayesian algorithm for detecting and correcting such faults.

Dharma P. Agrawal~\cite{AA} summarized many underlying design issues of WSNs, starting from the coverage and the connectivity. As batteries provide energy to sensor nodes, effective ways of power conservation are considered. Advantages of placing sensors in a regular pattern have also been discussed and various trade offs for many possible ways of secured communication in a WSN are summarized. Challenges in deploying WSN for monitoring emission are briefly covered. Finally, the use of sensors is illustrated in automatically generating music based on dancer's movements.

In almost all previous work, authors assume that event occurs over a region and there are fusion sensors that collect the information locally and take a decision. Since they do not introduce the concept of base station there is no concept of response probability. Also they assume the information are spatially correlated. Unlike the previous work, in this paper, we assume that there is no fusion sensor, all the sensors send the information to the base station and we introduce the probability model in two different stages, firstly, when a sensor detect the event and secondly when a sensor send the message to the base station. In the previous work, authors simulate the different probabilities for some specific values of parameters. In this paper, we calculate the exact probabilities and the exact test. It is hard to calculate the exact probabilities and the exact test for general case when events occur in more than one cell. In this paper, we assume that if event occurs then it occurs at only one cell.

In almost all previous work, authors assume that the grid as a square grid but the hexagonal grid is better in the sense that less number of sensors is required to cover the entire ROI. The minimum number of sensors needed to cover an area is given by Williams~\cite{W}. In this paper, we consider the both square and hexagonal grids in separate sections.

\section{Problem Statement and Notations}
In this section, we describe the problem that we want to solve and the notations we use.

\subsection{Problem Statement}
Our problem is to find various error probabilities (e.g., probability of false response when the ROI is normal or probability of no response when a particular square is a event square, etc.). We want to develop schemes for base station to take the decision and find the error probabilities of two different wrong situations: 1) base station decided that the ROI is normal but there is an event square and 2) base station decided that the ROI is not normal, i.e., a event square exists, but there is actually no event square. We want to develop the schemes and find the error probabilities under two different considerations: a) classical hypothesis testing and b) decision theoretic approach (i.e., Bayes test). In the consideration of decision theoretic approach, we introduce risk factor for two different wrong situations.  We also calculate error probabilities for some values of different parameters like probability of false alarm of a sensor, probability of event detection by a sensor, prior probability of existence of an event etc. Our problem is to give optimal test for base station (for different parameters) under the two different considerations and find some theoretical results.

Since we assume there is at most one particular event square and only $9$ sensors (one sensor which placed at the center of the event square, four sensor whose placed at the center of adjacent squares with a common side and four sensors which are placed at the center of adjacent squares with a common vertex) can only detect the event square, we consider a $3 \times 3$ square grid. Now if an event occurs, it occurs at the center square. Among the $9$ squares, our problem is to find whether or not the center square is the event square.

\begin{figure}[htb]
\begin{center}
\includegraphics[width = 7cm]{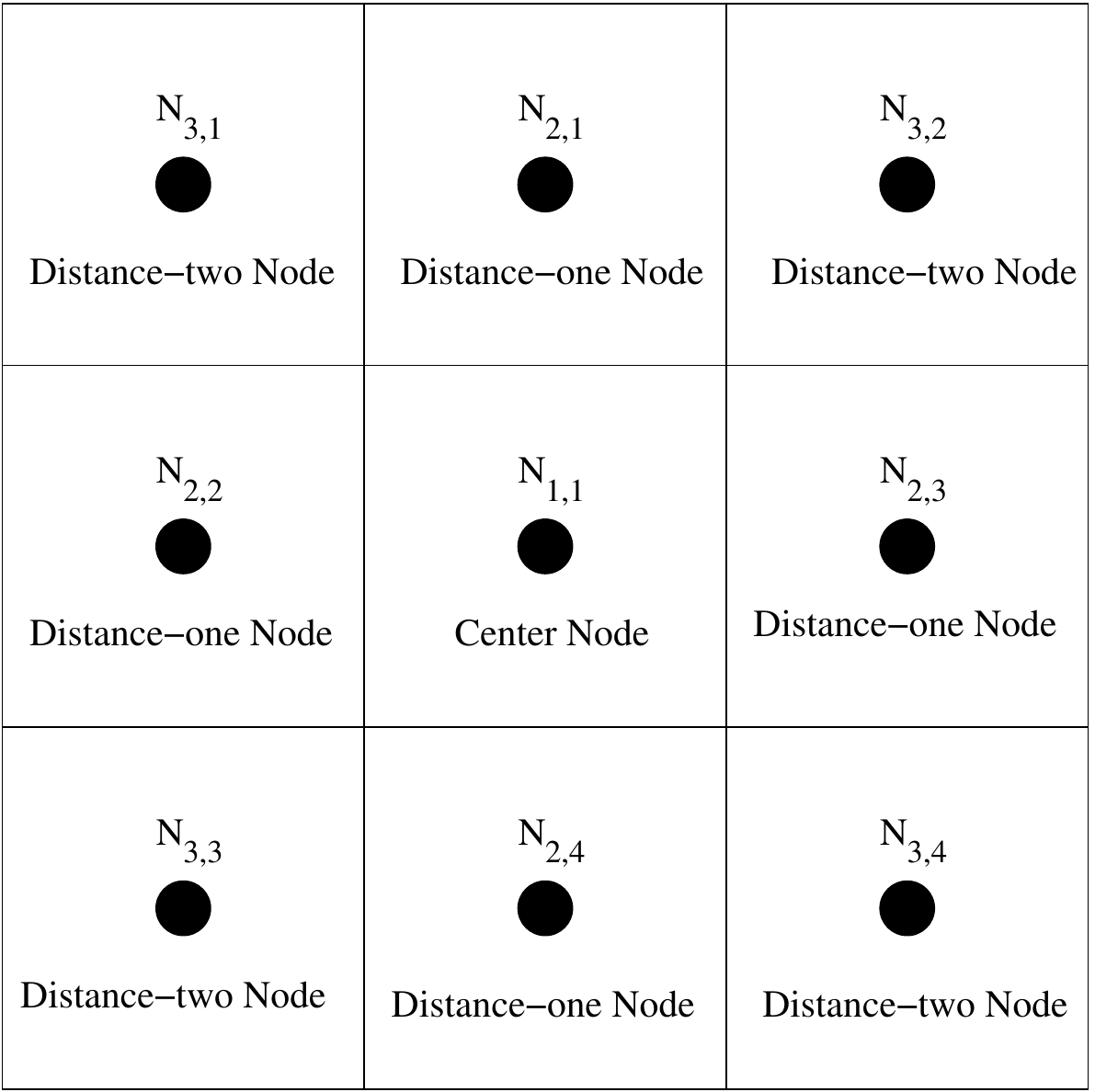}
\end{center}
\caption{Nodes placed in cells of the ROI}
\end{figure}

\subsection{Notations and Parameters}
The node which is placed at the center square is the nearest node and hence can detect the event square with highest probability. We denote this node as $N_{11}$.

The $4$ nodes, whose distances are $2a$ from the previous node, are the second nearest nodes and hence can detect the event square with second highest probability. We denote these nodes as $N_{2j}, j =1,2,3,4$.

The $4$ nodes, whose distances are $2 \sqrt{2} a$ from the center node, are the furthest nodes which can detect the event square and hence can detect the event square with lowest probability. We denote these nodes as $N_{3j}, j =1,2,3,4$.

For $(i,j)\in \{(1,1)\} \cup (\{2,3\} \times \{1,2,3,4\})$ \\
let $y_{ij}=1$ if the node $N_{ij}$ detects the center square as the event square,

    $y_{ij}=0$ if the node $N_{ij}$ detects the center square as the normal square (event does not occur),

    $x_{ij}=1$ if the node $N_{ij}$ responds, i.e., the node informs the base station that the center square is the event square, and

    $x_{ij}=0$ if the node $N_{ij}$ does not respond, i.e., the node informs the base station that the center square is normal.

Here we make one natural assumption, for $k,l=0,1$,
$$ \Pr(x_{ij}=k \ | \ y_{ij}=l, \Normal)= \Pr(x_{ij}=k \ | \ y_{ij}=l) $$
$$ \mbox{and} \ \Pr(x_{ij}=k \ | \ y_{ij}=l, \Event)= \Pr(x_{ij}=k \ | \ y_{ij}=l), $$
i.e., the response of a sensor is independent of the event occurrence and its ability to detect the event. 

Note that, detection of event by a sensor does not mean that the sensor informs the base station that the center square is the event square; if the sensor is faulty, it can send a normal report. Similar thing can happen if sensor does not detect the event square. Also note that, $y_{ij}$'s are not independent, but $y_{ij} $'s are independent under event and normal situation.

Now the parameters of the problem are the following:
\begin{itemize}
\item $\Pr(\Event) = \Pr(\mbox{event occurs}) = p_e \ \mbox{(say);}$
\item $\Pr(\Normal) = \Pr(\mbox{ROI is normal}) = p_n \ \mbox{(say)};$
\item $\Pr(y_{ij}=1 \ | \ \Event) = p_i \ \mbox{(say);}$
\item $\Pr(x_{ij}=1 \ | \ y_{ij}=1) = p_c \ \mbox{(say) and}$
\item $\Pr(x_{ij}=1 \ | \ y_{ij}=0) = p_w \ \mbox{(say)}$\\
for all possible values of $i$ and $j$.
\end{itemize}
Clearly, $\Pr(y_{ij}=1 \ | \ \Normal) = 0 $ for all possible values of $i$ and $j$. We also assume that $p_1>p_2>p_3$.

\section{Theoretical Analysis of Fault Detection}

In this section, we derive various error probabilities for all nodes and then propose a rule for the base station to take a decision compiling the information coming from all the $9$ nodes and to find the optimal solution. Finally, we calculate the error probabilities and the tests for the base station.

Let us consider the testing problem $H_0$: Event vs. $H_1$: Normal. We consider ``Event'' as null hypothesis because type I error should be the more serious error than type II error. If we reject the null hypothesis when it is true, i.e., if Event occurs but base station decides Normal, then that will be the more serious error than the other one.

There are two types of error: type I error when event occurs but sensor reports normal (which is the more serious error) and type II error when ROI is normal but sensor reports Event.

Throughout the section, we consider $i=1,j=1$ and $j=1,2,3,4$ when $i=2,3$. There are eight possible scenarios for a particular node $N_{i,j}$:

1. Normal, $y_{ij}=0, x_{ij}=0$ (sensor correctly detects a normal reading and sends the correct message to the base station),

2. Normal, $y_{ij}=0, x_{ij}=1$ (sensor correctly detects a normal reading but sends the wrong message to the base station due to fault),

3. Normal, $y_{ij}=1, x_{ij}=0$ (sensor wrongly detects a normal reading as event but sends the normal message to the base station due to fault),

4. Normal, $y_{ij}=1, x_{ij}=1$ (sensor wrongly detects a normal reading as event and sends the wrong message to the base station),

5. Event, $y_{ij}=0, x_{ij}=0$ (sensor wrongly detects an event reading and sends the wrong message i.e. normal message to the base station),

6. Event, $y_{ij}=0, x_{ij}=1$ (sensor wrongly detects an event reading but sends the correct message to the base station),

7. Event, $y_{ij}=1, x_{ij}=0$ (sensor correctly detects an event reading but sends the wrong message to the base station due to fault), and

8. Event, $y_{ij}=1, x_{ij}=1$ (sensor correctly detects an event reading and sends the correct message to the base station).

\tikzstyle{level 1}=[level distance=2.5cm, sibling distance=3.5cm]
\tikzstyle{level 2}=[level distance=2.5cm, sibling distance=2cm]

\tikzstyle{bag} = [text width=4em, text centered]
\tikzstyle{end} = [circle, minimum width=3pt,fill, inner sep=0pt]

\begin{figure}
\begin{tikzpicture}[grow=right, sloped]

\node[bag] {Normal}
    child {
        node[bag] {$ y_{ij}=1 $}
            child {
                node[end, label=right:
                    {$x_{ij}=1 $}] {}
                edge from parent
                node[above] {$ $}
                node[below]  {$ $}
            }
            child {
                node[end, label=right:
                    {$x_{ij}=0 $}] {}
                edge from parent
                node[above] {$ $}
                node[below]  {$ $}
            }
            edge from parent
            node[above] { detection}
            node[below]  {$0 $}
    }
    child {
        node[bag] { $ y_{ij}=0 $}
        child {
                node[end, label=right:
                    {$x_{ij}=1 $}] {}
                edge from parent
                node[above] {response}
                node[below]  {$p_w $}
            }
            child {
                node[end, label=right:
                    {$x_{ij}=0 $}] {}
                edge from parent
                node[above] {$1-p_w $}
                node[below]  {$ $}
            }
        edge from parent
            node[above] {$1$}
            node[below]  {$ $}
    };

\end{tikzpicture}
\caption{detection and response probabilities when ROI is normal}
\end{figure}
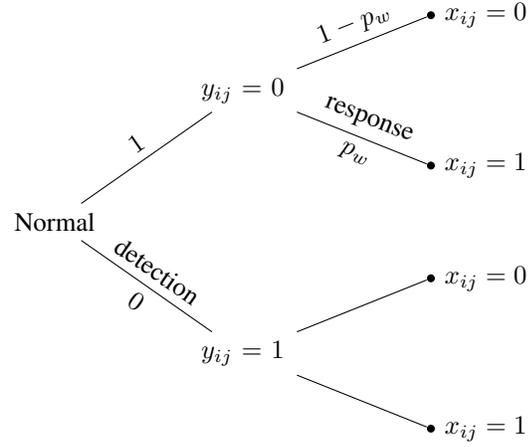

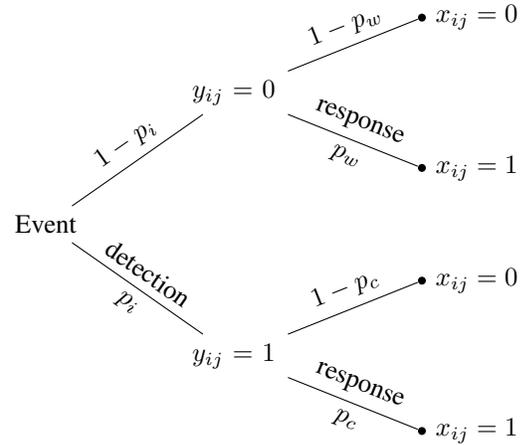
\begin{figure}
\begin{tikzpicture}[grow=right, sloped]

\node[bag] {Event}
    child {
        node[bag] {$ y_{ij}=1 $}
            child {
                node[end, label=right:
                    {$x_{ij}=1 $}] {}
                edge from parent
                node[above] {response}
                node[below]  {$p_c$}
            }
            child {
                node[end, label=right:
                    {$x_{ij}=0 $}] {}
                edge from parent
                node[above] {$1-p_c$}
                node[below]  {$ $}
            }
            edge from parent
            node[above] {detection}
            node[below]  {$p_i $}
    }
    child {
        node[bag] { $ y_{ij}=0 $}
        child {
                node[end, label=right:
                    {$x_{ij}=1 $}] {}
                edge from parent
                node[above] {response}
                node[below]  {$p_w $}
            }
            child {
                node[end, label=right:
                    {$x_{ij}=0 $}] {}
                edge from parent
                node[above] {$1-p_w $}
                node[below]  {$ $}
            }
        edge from parent
            node[above] {$1-p_i$}
            node[below]  {$ $}
    };
\end{tikzpicture}
\caption{detection and response probabilities when event occurs}
\end{figure}

\subsection{Error Probabilities for Nodes}
\vspace{-0.5cm}
\begin{eqnarray*}
&& \mbox{Let,\ }P_N = \Pr(x_{ij}=0 \ | \ \Normal)\\
&& = \Pr(x_{ij}=0 \ | \ y_{ij}=0, \Normal)\Pr(y_{ij}=0 \ | \ \Normal) \\
&& +  \Pr(x_{ij}=0 \ | \ y_{ij}=1, \Normal)\Pr(y_{ij}=1 \ | \ \Normal)\\
&& = \Pr(x_{ij}=0 \ | \ y_{ij}=0)\Pr(y_{ij}=0 \ | \  \Normal) \ + \\
&&\Pr(x_{ij}=0 \ | \ y_{ij}=1)\Pr(y_{ij}=1 \ | \  \Normal) = 1 - p_w.\\
\end{eqnarray*}
Therefore, the probability of type II error for the node $(i,j)$ is $\Pr(x_{ij}=1 \ | \ \Normal) = p_w$
\begin{eqnarray*}
&& \mbox{Let, \ } P_{E,i} = \Pr(x_{ij}=1 \ | \ \Event) = \\
&&\Pr(x_{ij}=1 \ | \ y_{ij}=0, \Event)\Pr(y_{ij}=0 \ | \ \Event) + \\
&&\Pr(x_{ij}=1 \ | \ y_{ij}=1, \Event)\Pr(y_{ij}=1 \ | \ \Event) \\
&&= \Pr(x_{ij}=1 \ | \ y_{ij}=0)\Pr(y_{ij}=0 \ | \ \Event) + \\
&&\Pr(x_{ij}=1 \ | \ y_{ij}=1)\Pr(y_{ij}=1 \ | \ \Event) \\
&&=  p_w(1-p_i) + p_c p_i = p_w + p_i (p_c - p_w)\\
\end{eqnarray*}
Hence, probability of type I error for the node $ (i,j)$ is
$$Q_{E,i} = \Pr(x_{ij}=0 \ | \ \Event) = 1 -\Pr(x_{ij}=1 \ | \ \Event) $$
$\ \ = (1-p_w) - p_i ( p_c - p_w)$

Now other types of errors may be as follows:
$$P_{1,i} = \Pr(\Event  \ | \ x_{ij}=0) = \frac {p_e Q_{E,i}}{p_n P_N + p_e Q_{E,i}}$$
$$\mbox{and}\ P_{2,i} = \Pr(\Normal  \ | \ x_{ij}=1) =  \frac {p_n p_w}{p_n p_w + p_e P_{E,i}}.$$

\subsection{Error Probabilities and Test for Base Station}

Now, let us consider the detection problem for the base station. After the observations about $x_{ij}$'s are made, at the base station they are combined to make a final decision regarding the hypotheses ($H_0$: Event vs $H_1$: Normal). When $H_0$ is true, $x_{ij}$ follows $\Ber(P_{E,i})$, and when $H_1$ is true $x_{ij}$ follows $\Ber(p_w)$. Let the probability mass function of $x_{ij}$ when $H_k$ is true be $p(x_{ij}  \ | \  H_k)$ for $k=0,1$. We make one more natural assumption $p_c >p_w$, which is equivalent to say $P_{E,i} > p_w$ for all $i$. This is needed for a result in the next section .

\subsubsection{The Neyman-Pearson Approach}

In many practical situations, the prior probabilities may be unknown in which case the decision theoretic approach is not appropriate. So, we employ the Neyman-Pearson criterion. In that case, the most powerful (MP) test of size $\a$ is to reject $H_0$ when
$$\Pi p(x_{ij}  \ | \  H_1) > \l'' \Pi p(x_{ij}  \ | \  H_0) $$
and reject $H_0$, with probability $k$, when
$$\Pi p(x_{ij}  \ | \  H_1) = \l'' \Pi p(x_{ij}  \ | \  H_0), $$
where $\l''$ and $k$ can be found from the size $\a$ of the test. Since, when $H_0$ is true $x_{ij}$ follows $\Ber(P_{E,i})$ and, when $H_1$ is true, $x_{ij}$ follows Ber($p_w$), we can simplify the MP test as to reject $H_0$ when
$$p_w^{\Si x_{ij}} (1- p_w)^{\Si(1-x_{ij})} > \l'' P_{E,1} ^{x_{11}} (1-P_{E,1})^{(1-x_{11})} \times $$
$$ P_{E,2} ^{\Si x_{2j}} (1-P_{E,2})^{\Si(1-x_{2j})}P_{E,3} ^{\Si x_{3j}} (1-P_{E,3})^{\Si(1-x_{3j})}, $$
and reject $H_0$ with probability $k$ when
$$p_w ^{\Si x_{ij}} (1- p_w)^{\Si(1-x_{ij})} = \l'' P_{E,1} ^{x_{11}} (1-P_{E,1})^{(1-x_{11})} \times $$
$$P_{E,2} ^{\Si x_{2j}} (1-P_{E,2})^{\Si(1-x_{2j})} P_{E,3} ^{\Si x_{3j}} (1-P_{E,3})^{\Si(1-x_{3j})}.$$
i.e., reject $H_0$ when
$$ \frac{1}{ \l''} >\left(\frac{P_{E,1}}{p_w}\right) ^{x_{11}} \left(\frac{1-P_{E,1}}{1- p_w}\right)^{(1- x_{11})} \left(\frac{P_{E,2}}{p_w}\right) ^{\Si x_{2j}} \times$$
$$\left(\frac{1-P_{E,2}}{1- p_w}\right)^{(4-\Si x_{2j})}\left(\frac{P_{E,3}}{p_w}\right) ^{\Si x_{3j}} \left(\frac{1-P_{E,3}}{1- p_w}\right)^{(4-\Si x_{3j})}$$
and  reject $H_0$ with probability k when equality hold in place of greater than.\\
Hence, reject $H_0$ when
$${x_{11}} ln\left(\frac{P_{E,1}}{p_w}\right) + (1-x_{11}) ln\left(\frac{1-P_{E,1}}{1-p_w}\right) + $$
$${\Si x_{2j}} ln\left(\frac{P_{E,2}}{Q_N}\right) + (4-\Si x_{2j}) ln\left(\frac{1-P_{E,2}}{1- p_w}\right) + $$
$${\Si x_{3j}} ln\left(\frac{P_{E,3}}{p_w}\right) + (4-\Si x_{3j}) ln\left(\frac{1-P_{E,3}}{1- p_w}\right) < \l'$$
and reject $H_0$ with probability k when equality hold in place of less than.\\
i.e., reject $H_0$ when
$${x_{11}} ln\left(\frac{P_{E,1}(1- p_w)}{(1-P_{E,1})p_w}\right) + {\Si x_{2j}}ln\left(\frac{P_{E,2}(1- p_w)}{(1-P_{E,2})p_w}\right) + $$
$${\Si x_{3j}}ln\left(\frac{P_{E,3}(1- p_w)}{(1-P_{E,3})p_w}\right) < \l$$
and reject $H_0$ with probability k when equality hold in place of less than.\\
Hence, we get the MP test as to reject $H_0$ when
$${\Si x_{ij}} ln\left(\frac{P_{E,i}(1-p_w)}{(1-P_{E,i})p_w}\right) < \l$$
and reject $H_0$  with probability $k$ when
$${\Si x_{ij}} ln\left(\frac{P_{E,i}(1- p_w)}{(1-P_{E,i})p_w}\right) = \l \mbox{\ \ ...\ (R)}$$
where, $\l$ and $k$ can be found from the relation
$$\Pr(H_0 \mbox{\ reject} \ | \ H_0 \mbox{\ true}) =  \a, $$
which is equivalent to
$$\Pr({\Si x_{ij}} ln(d_i) < \l ) + k\Pr({\Si x_{ij}} ln(d_i) = \l ) = \a,$$
where, $d_i = \frac{P_{E,i}(1- p_w)}{(1-P_{E,i})p_w}$ and $x_{ij}$ follows $\Ber(P_{E,i})$. Since we assume $P_{E,i} > p_w$, $ln\left(\frac{P_{E,i}(1- p_w)}{(1-P_{E,i})p_w}\right)>0 $ for all $i$.

Based on the given error bound $ \a $ and sensor fault probabilities, the base station will take the decision given by the rule (R).

\subsubsection{Decision Theoretic Approach}

A test $T_g$ of $H_0: \theta = \theta_0 $ vs $H_1: \theta = \theta_1 $ is defined to be a Bayes test with respect to the prior distribution $\Pr(H_1) = g$ if and only if
$$(1-g)R_{T_g}(\theta_0) + gR_{T_g}(\theta_1) \leq (1-g)R_T(\theta_0) + gR_T(\theta_1)$$
for any other test $T$, where $R_T(\theta)$ is the risk function of the test $T$. The Bayes test is the test which seeks a critical region that minimizes the overall risk. If loss function is not available then we can assume the losses are $0$ or $1$. It can be proved that the Bayes test is to Reject $H_0$ when
$$\frac{L_0}{L_1} < \frac{gl(d_0;\theta_1)}{(1-g)l(d_1;\theta_0)},$$
where $L_0$ and $L_1$ are the likelihoods for $\theta = \theta_0 $ and $ \theta = \theta_1$, respectively; $l(d_0;\theta_1)$ is the loss when null hypothesis is accepted but it is false, and $l(d_1;\theta_0)$ is the loss when null hypothesis is rejected but it is true \cite{MGB}.

Let the losses be as follows: $l_e$ when event occurs but base station takes decision as normal, $l_n$ when ROI is normal but base station takes decision as event, and loss is $0$ when the base station takes the correct decision. Hence, under the Bayesian setup, i.e., when the prior distribution $(p_n,p_e)$ are available, the Bayes test with respect to the prior distribution $\Pr(H_0) = p_e$ and $\Pr(H_1) = p_n$ can be derived as follows:

Reject $H_0$ when
$$p_w ^{\Si x_{ij}} (1- p_w)^{\Si(1-x_{ij})} >  \frac{p_e l_e}{p_n l_n} P_{E,1} ^{x_{1j}} (1-P_{E,1})^{(1-x_{1j})} \times $$
$$P_{E,2} ^{\Si x_{2j}} (1-P_{E,2})^{\Si(1-x_{2j})} P_{E,3} ^{\Si x_{3j}} (1-P_{E,3})^{\Si(1-x_{3j})},$$

i.e., Reject $H_0$ when
$${\Si x_{ij}} ln\left(\frac{P_{E,i}(1-p_w)}{(1-P_{E,i})p_w}\right) < ln\left(\frac{p_n l_n}{p_e l_e}\right) + ln\left(\frac{1-p_w}{1-P_{E,1}}\right)$$
$$ + \ 4ln\left(\frac{1-p_w}{1-P_{E,2}}\right) + 4ln\left(\frac{1-p_w}{1-P_{E,3}}\right)$$

\subsection{Boundary Case}

In the above discussion, we assume that the center square is an interior one. If we consider the boundary squares, then the expression for the error probabilities for each sensor are changed and consequently expression for the error probabilities for the base station are also changed. In that case, if we consider the corner squares then $j$ takes value $1$ for $i=1,3$ and $j$ takes values $1,2$ for $i=2$, and if we consider the boundary square other than a corner one then $j$ takes value $1$ for $i=1$; $ j $ takes values $1, 2,3 $ for $i=2$, and $j$ takes values $1, 2$ for $i=3$. The theoretical analysis is similar as in the case of interior squares.

\subsection{When More Sensors can Detect the Event Square}

We may consider the situation when sensing radius has larger value, and then more sensors can detect the event square but with different probabilities. In this case, we classify all the nodes as follows: two sensors belong to the same class if they have the same distance from the event square and hence the same detection probability to detect the event. Let sensors in the $i$-th class detect the event square with probability $p_i, i=1,2,3,...$. Then, the expressions of the error probabilities for the sensors and the test (MP and Bayes) are similar to the ones previously discussed, but now the summation in the left hand side of the expression of MP or Bayes test is changed; instead of three terms, there will be more terms.

\section{Calculations and Observations}

We have the independent set of parameters of the problem as follows:

$p_e$, $p_i$, $p_c$, $p_w$  for all possible values of $i$ and $j$,

$l$ = ratio of losses = $l_e$/$l_n$ and size of the test = $\a$.

The type I error for $N_{ij}$ is $ (1-p_w) - p_i ( p_c - p_w)$
and the type II error for $N_{ij}$ is $p_w$ for all possible values of $i$ and $j$.

Let $d = p_c - p_w$ and $d>0$.

Other types of errors for $N_{ij}$'s are
$$P_{1,i} = \frac{p_e(1-p_w-p_id)}{1-p_w-p_ep_id} \mbox{ \ and \ } P_{2,i} = \frac{(1-p_e)p_w}{p_w+p_ep_id}$$
for all possible values of $i$ and $j$,

$$\mbox{Let \ } t_i = ln\left(\frac{P_{E,i}(1- p_w)}{(1-P_{E,i})p_w}\right).$$
$$\mbox{Therefore, \ }  t_i = ln\left(\frac{(p_w+p_id)(1- p_w)}{p_w(1-p_w-p_id)}\right)$$
$$ = ln\left(1 + \frac{ p_id}{p_w(1-p_w-p_id)}\right).$$

Let $t = ln\left(\frac{p_n l_n}{p_e l_e}\right) + $
$$ln\left(\frac{1-p_w}{1-P_{E,1}}\right) + 4ln\left(\frac{1-p_w}{1-P_{E,2}}\right) + 4ln\left(\frac{1-p_w}{1-P_{E,3}}\right)$$
$$= ln\left(\frac{1-p_e}{lp_e}\right) + ln\left(\frac{1-p_w}{1-p_w - p_1d}\right) + $$
$$4ln\left(\frac{1-p_w}{1-p_w - p_2d}\right) + 4ln\left(\frac{1-p_w}{1-p_w - p_3d}\right)$$
$$= ln\left(\frac{1-p_e}{lp_e}\right) + ln\left(1 + \frac{p_1d}{1-p_w - p_1d}\right) +  $$
$$4ln\left(1 + \frac{p_2d}{1-p_w - p_2d}\right) + 4ln\left(1 + \frac{p_3d}{1-p_w - p_3d}\right).$$

Also let $x_1= x_{11}$ and $x_i= x_{i1}+ x_{i2}+ x_{i3}+ x_{i4}$ for $i=2,3.$

\subsection{Calculation of Errors for Each Sensor and Observations}

In this subsection, we calculate different error probabilities for sensors for some specific values of parameters. We choose two set of values of $p_1, p_2, p_3, p_w$ and $p_c$, one is for a good reliable network and other is for a less reliable network. We choose five different values of $p_e$. These values are chosen just to give an idea of the errors and the tests. One can easily calculate different error probabilities for any other values of the parameters.

\begin{table}[htb]
\caption{Calculation of errors for some values of the parameters}
{
\begin{center}
\begin{tabular}{|c|c|c|c|c|c|c|c|}
\hline
\multicolumn{8}{|c|} {Calculation of type I error} \\ \hline \hline
    $p_1$ & $p_2$ & $p_3$ & $p_c$ & $p_w$ & $Q_{E,1}$ & $Q_{E,2}$ & $Q_{E,3}$\\ \hline
    0.9 & 0.5 & 0.3 & 0.9 & 0.1 & 0.1800 & 0.5000 & 0.6600 \\ \hline
    0.7 & 0.3 & 0.1 & 0.8 & 0.2 & 0.3800 & 0.6200 & 0.7400  \\ \hline \hline
\end{tabular}
\begin{tabular}{|c|c|c|c|c|c|c|}
\hline
\multicolumn{7}{|c|} {When $p_1 = 0.9,p_2 = 0.5,p_3 = 0.3, p_w = 0.1, p_c = 0.9$} \\ \hline \hline
    $p_e$ & $P_{1,1}$ & $P_{1,2}$ & $P_{1,3}$ & $P_{2,1}$ & $P_{2,2}$ & $P_{2,3}$  \\ \hline
    0.1 & 0.0217 & 0.0581 & 0.0753 & 0.5233 & 0.6429 & 0.7258 \\ \hline
    0.2 & 0.0476 & 0.1220 & 0.1550 & 0.3279 & 0.4444 & 0.5405 \\ \hline
    0.3 & 0.0789 & 0.1923 & 0.2391 & 0.2215 & 0.3182 & 0.4070 \\ \hline
    0.4 & 0.1176 & 0.2702 & 0.3284 & 0.1546 & 0.2308 & 0.3061 \\ \hline
    0.5 & 0.1667 & 0.3571 & 0.4231 & 0.1087 & 0.1667 & 0.2273 \\ \hline \hline
\end{tabular}
\begin{tabular}{|c|c|c|c|c|c|c|}
\hline
\multicolumn{7}{|c|}{When $p_1 = 0.7,p_2 = 0.3,p_3 = 0.1, p_w = 0.2, p_c = 0.8$} \\ \hline \hline
    0.1 & 0.0501 & 0.0793 & 0.0932 & 0.7438 & 0.8257 & 0.8738  \\ \hline
    0.2 & 0.1061 & 0.1623 & 0.1878 & 0.5634 & 0.6780 & 0.7547  \\ \hline
    0.3 & 0.1691 & 0.2493 & 0.2839 & 0.4294 & 0.5511 & 0.6422  \\ \hline
    0.4 & 0.2405 & 0.3407 & 0.3814 & 0.3261 & 0.4412 & 0.5357  \\ \hline
    0.5 & 0.3220 & 0.4366 & 0.4805 & 0.2439 & 0.3448 & 0.4348  \\ \hline

\end{tabular}
\end{center}
}

\label{simtable 1}
\end{table}

{\it Observations:} A few immediate observations from the theoretical results (which can be verified from Table~\ref{simtable 1}) are as follows:

\begin{enumerate}
\item $ P_{1,i} $ decreases when $p_i$ and $ p_c$ increase and independent of $p_j$ when $i\neq j$.
\item $ P_{1,i} $ increases when $p_e$ and $ p_w$ increase.
\item $ P_{2,i} $ decreases when $p_i$, $p_e$ and $ p_c$ increase and independent of $p_j$ when $i\neq j$.
\item $ P_{2,i} $ increases when $ p_w$ increases.
\item When $p_e$ and $p_w$ are small then
\begin{eqnarray*}
&& P_{1,i} =  \frac{p_e(1-p_w-p_id)}{1-p_w-p_ep_id} \\
&& =p_e\left(1-\frac{p_id}{1-p_w}\right)\left(1- \frac{p_ep_id}{1-p_w}\right)^{-1} \\
&& \approx p_e\left(1-\frac{p_id}{1-p_w}\right)\left(1+ \frac{p_ep_id}{1-p_w}\right) \\
&& \approx p_e\left(1-\frac{p_id}{1-p_w} + \frac{p_ep_id}{1-p_w}\right) \\
\end{eqnarray*}
\item If detection probability $p_i$ is low then type I error is close to $1$. In that case, the network is not reliable.
\item If $p_e$ is high but $p_i$ is small then other types of errors are high; that means, there is no use of sensors. So when $p_e$ is high, we have to use sensors with high detection probability (i.e. better quality sensors).

\end{enumerate}

\subsection{Calculation for Bayes Test and Observations}

The Bayes test is to reject $H_0$ when
$$\l_1x_1 + \l_2x_2 + \l_3x_3 < 1,$$
where, $\l_i = t_i/t $ with
$$t_i = ln\left(1+ \frac{p_id}{p_w(1-p_w-p_id)}\right), i=1,2,3.$$
$$\mbox{and} \ t = ln\left(\frac{1-p_e}{lp_e}\right) + ln\left(1 + \frac{p_1d}{1-p_w - p_1d}\right) + $$
$$4ln\left(1 + \frac{p_2d}{1-p_w - p_2d}\right) + 4ln\left(1 +\frac{p_3d}{1-p_w - p_3d}\right)$$

Note that we always accept $H_0$ if $t \leq 0$
$$\mbox{As \ } ln\left(1 + \frac{p_id}{1-p_w - p_id}\right)>0, \mbox{\ for all\ } i,$$
$t$ is negative if
$$ln\left(\frac{lp_e}{1-p_e}\right) > ln\left(1 + \frac{p_1d}{1-p_w - p_1d}\right)\ + $$
$$4ln\left(1 + \frac{p_2d}{1-p_w - p_2d}\right) + 4ln\left(1 +\frac{p_3d}{1-p_w - p_3d}\right),$$
$$\mbox{i.e., if \ } l> \left(\frac{1-p_e}{p_e}\right)\left(1 + \frac{p_1d}{1-p_w - p_1d}\right)\times$$
$$\left(1 + \frac{p_2d}{1-p_w - p_2d}\right)^4\left(1 +\frac{p_3d}{1-p_w - p_3d}\right)^4.$$
In this case, the Bayes test is not applicable, i.e., if $l$ (ratio of the losses) is large then we have to use good quality sensors, i.e., sensor with high detection probabilities such that $\frac{p_1d}{1-p_w - p_1d}$ is so large that
$$l< \left(\frac{1-p_e}{p_e}\right)\left(1 + \frac{p_1d}{1-p_w - p_1d}\right)$$
$$\left(1 + \frac{p_2d}{1-p_w - p_2d}\right)^4\left(1 +\frac{p_3d}{1-p_w - p_3d}\right)^4 .$$

\begin{table}[htb]
\caption{Calculation of Bayes test for some values of the parameters}
{
\begin{center}
\begin{tabular}{|c|c|c|c|c|c|c|c|}
\hline
\multicolumn{4}{|c|}{$p_1 = 0.9,p_2 = 0.5,p_3 = 0.3, p_w = 0.1, p_c = 0.9$} \\ \hline \hline
    $p_e$ & $l$ & $t$ & Bayes test \\ \hline
    0.1 & 5 & 5.789 & $3.714x_1 + 2.197x_2 + 1.534x_3 \leq 5.789 $    \\ \hline
    0.3 & 5 & 4.439 & $3.714x_1 + 2.197x_2 + 1.534x_3 \leq 4.439 $    \\ \hline
    0.5 & 5 & 3.592 & $3.714x_1 + 2.197x_2 + 1.534x_3 \leq 3.592 $    \\ \hline
    0.1 & 20 &4.403 & $3.714x_1 + 2.197x_2 + 1.534x_3 \leq 4.403 $  \\ \hline
    0.3 & 20 &3.053 & $3.714x_1 + 2.197x_2 + 1.534x_3 \leq 3.053 $  \\ \hline
    0.5 & 20 &2.205 & $3.714x_1 + 2.197x_2 + 1.534x_3 \leq 2.205 $  \\ \hline \hline
\multicolumn{4}{|c|}{$p_1 = 0.7,p_2 = 0.3,p_3 = 0.1, p_w = 0.2, p_c = 0.8$} \\ \hline \hline
    0.1 & 5 & 2.664 & $1.876x_1 + 0.897x_2 + 0.340x_3 \leq 2.664 $    \\ \hline
    0.3 & 5 & 1.314 & $1.876x_1 + 0.897x_2 + 0.340x_3 \leq 1.314 $     \\ \hline
    0.5 & 5 & 0.466 & $1.876x_1 + 0.897x_2 + 0.340x_3 \leq 0.466 $     \\ \hline
    0.1 & 20 &1.277 & $1.876x_1 + 0.897x_2 + 0.340x_3 \leq 1.277 $    \\ \hline
    0.3 & 20 & -0.073 & $1.876x_1 + 0.897x_2 + 0.340x_3 \leq -0.073 $    \\ \hline
    0.5 & 20 & -0.920 & $1.876x_1 + 0.897x_2 + 0.340x_3 \leq -0.920 $    \\ \hline
\end{tabular}
\end{center}
}

\label{simtable 2}
\end{table}

{\it Observations:} A few immediate observations from the theoretical results (which can be verified from Table~\ref{simtable 2}) are as follows:

\begin{enumerate}
\item $\l_i$'s are the weights of $x_i$'s in the Bayes test which means that the value of $\l_i$ tell us how much weight the base station has to give to $x_i$ while taking the decision about the event square, e.g., consider the Bayes test for $p_1=0.9,p_2=0.5,p_3=0.3,p_w=0.1,p_c=0.9,p_e=0.1$ and $l=5$ (Table ~\ref{simtable 2}),
$$3.714x_1 + 2.197x_2 + 1.534x_3 \leq 5.789,$$
Here, $\l_1:\l_2:\l_3 \approx 5:3:2$ means $3$ distance-two sensors is equivalent to $2$ distance-one sensors in the context of detecting an event and so on.
\item Let $\l_i / \l_k$= $t_i / t_k$ be the ratio of the weights which tells us how many $N_{kj}$ nodes are equivalent to one $N_{ij}$ in the context of detecting an event.
\item $t$ increases when $p_i$ and $ p_c$ increase.
\item $t$ decreases when $p_w, l$ and $ p_e$ increase.
\item For $p_1 = 0.7,p_2 = 0.3,p_3 = 0.1, p_w = 0.2, p_c = 0.8, l = 20$ and $p_e = 0.3$ (resp. 0.5), the Bayes test is (Table~\ref{simtable 2}) to reject $H_0$ when
$1.876x_1 + 0.897x_2 + 0.340x_3 \leq -0.0734$
$$(\mbox{resp.\ } 1.876x_1 + 0.897x_2 + 0.340x_3 \leq -0.920  ),$$
i.e., we accept $H_0$ for all values of $x_i$'s. So in this situation the Bayes test is not applicable. This indicates that for small values of $p_i $'s and large values of $ l $ the Bayes test is not applicable. When $l$ is large we have to use sensors with high detection probabilities to decide about the event square using Bayes test.

\end{enumerate}

\subsection{Calculation for Most Powerful Test and Observations}

The most powerful (MP) test of size $\a$ is to reject $H_0$ when
$$t_1x_1 + t_2x_2 + t_3x_3 < \l$$
and reject $H_0$ with probability $k$, when
$$t_1x_1 + t_2x_2 + t_3x_3 = \l,$$
$$\mbox{where,\ } t_i = ln\left(1+ \frac{p_id}{p_w(1-p_w-p_id)}\right), i=1,2,3.$$

$\l$ and $k$ can be found from the relation
\begin{eqnarray*}
&&\Pr(t_1x_1 + t_2x_2 + t_3x_3 < \l ) + \\
&& k\Pr(t_1x_1 + t_2x_2 + t_3x_3 = \l ) = \a.
\end{eqnarray*}
Note that, $x_1$ follows $\Ber(p_w + p_1d)$ and $x_i$ follows $\Bin(4,p_w + p_id)$ for $i=2,3$.\\
Also $x_i$'s are independent when it is known that $H_0$ is true.

To simplify calculations, we take the approximate values of $t_1: t_2 :t_3 $ and $p_w +p_id$.
For $p_1 = 0.9,p_2 = 0.5,p_3 = 0.3, p_w = 0.1, p_c = 0.9,$ we take $t_1: t_2 :t_3 \approx 5:3:2$ and $p_w +p_1d \approx 0.8, p_w +p_2d \approx 0.5$ and $p_w +p_3d \approx 0.35$.
And for $p_1 = 0.7,p_2 = 0.3,p_3 = 0.1, p_w = 0.2, p_c = 0.8,$ we take $t_1: t_2 :t_3 \approx 10:5:2$ and $p_w +p_1d \approx 0.6, p_w +p_2d \approx 0.4$ and $p_w +p_3d \approx 0.25$.

To calculate $\l$ and $k$, we first set $\l=0 $ and calculate the probability of
$$t_1x_1 + t_2x_2 + t_3x_3 < \l,$$
if the probability is less than $\a$, we increase the value of $\l$ by 1 and do the same as above. If, for $\l=\l'$ the probability is less than $\a$, and for $\l > \l'$ the probability is greater than $\a$, we take that $\l'$ as the value of $\l$, and then, calculate the value of $k$ as
$$k = \frac{\a-\Pr(X < \l')}{\Pr(X \leq \l')-\Pr(X < \l')},$$
where, $ X=t_1x_1 + t_2x_2 + t_3x_3$.
\begin{table}[htb]
\caption{Calculation of MP test for some values of the parameters (approximated)}
{
\begin{center}
\begin{tabular}{|c|c|c|c|c|c|c|c|c|}
\hline
\multicolumn{4}{|c|}{$p_1 = 0.9,p_2 = 0.5,p_3 = 0.3, p_w = 0.1, p_c = 0.9$} \\ \hline \hline
     $\a$ & $\l$ & $k$ & MP test \\ \hline
     0.900 & 8 & 0.04 & $5x_1 + 3x_2 + 2x_3 \leq 8$  \\ \hline
     0.950 & 6 & 0.15 & $5x_1 + 3x_2 + 2x_3 \leq 6$ \\ \hline
     0.975 & 5 & 0.19 & $5x_1 + 3x_2 + 2x_3 \leq 5$ \\ \hline
     0.990 & 3 & 0.33 & $5x_1 + 3x_2 + 2x_3 \leq 3$ \\ \hline\hline
\multicolumn{4}{|c|}{$p_1 = 0.7,p_2 = 0.3,p_3 = 0.1, p_w = 0.2, p_c = 0.8$} \\ \hline \hline
     $\a$ & $\l$ & $k$ & MP test \\ \hline
     0.900 & 7 & 0.24 & $10x_1 + 5x_2 + 2x_3 \leq 7 $  \\ \hline
     0.950 & 5 & 0.02 & $10x_1 + 5x_2 + 2x_3 \leq 5 $  \\ \hline
     0.975 & 2 & 0.39 & $10x_1 + 5x_2 + 2x_3 \leq 2 $  \\ \hline
     0.990 & 0 & 0.61 & $10x_1 + 5x_2 + 2x_3 \leq 0 $  \\ \hline

\end{tabular}
\end{center}
}

\label{simtable 3}
\end{table}

{\it Observations:} A few immediate observations from the theoretical results (which can be verified from  Table~\ref{simtable 3}) are as follows:

\begin{enumerate}
\item $t_i$'s are the weights of the $x_i$'s in the MP test
$$5x_1 + 3x_2 + 2x_3 \leq 8, \mbox{here \ } \l_1:\l_2:\l_3 \approx 5:3:2,$$
roughly means $3$ distance-two sensors is equivalent to $2$ distance-one sensors in the context of detecting event and so on.
\item $t_i$ is independent of $p_e$ and $l$.
\item $t_i$ increases when $p_i$ and $ p_c$ increase.
\item $t_i$ decreases when $p_w$ increases.
\item As $t_i$ increases, critical region (set of all the values of $x_i$'s for which we reject the null hypothesis) is going to be smaller.
\item Let $\l_i / \l_k$= $t_i / t_k$ be the ratio of the weights which tells us how many $N_{kj}$ node are equivalent to one $N_{ij}$ in the context of detecting an event.
\item For $p_1 = 0.7,p_2 = 0.3,p_3 = 0.1, p_w = 0.2, p_c = 0.8 $ and $ \a = 0.990 $ the MP test is reject $H_0$ when
$$10x_1 + 5x_2 + 2x_3 \leq 0,$$
i.e., we accept $H_0$ in almost all cases. So in this situation, the MP test is not applicable. This indicates that for small values of $ p_i $'s and large values of $ \a $, the MP test is not applicable. When $\a $ is large we cannot use sensors with small $ p_i $ values for the MP test. Hence, when the size of the MP test is small we have to use good enough sensors for MP test.
\end{enumerate}

\section{Simulation Results}
For $m=200, n=250$, we simulate the different probabilities and the tests of the problem. We also simulate the number of times the Bayes test and the MP test give the correct decision. The simulation is performed using the C-program with required random numbers generated using the standard C-library.

In the following table, $q_{E,i}$ denotes the corresponding simulated values of $Q_{E,i}$ and $p_{k,i}$ denotes the corresponding simulated values of $P_{k,i}$ where $k= 1,2$ and $i = 1, 2, 3$.

\begin{table}[htb]
\caption{Simulated and theoretical values of errors}
{
\begin{center}
\begin{tabular}{|c|c|c|c|c|c|}
\hline
\multicolumn{6}{|c|} {Simulation of type I error} \\ \hline \hline
    $Q_{E,1}$ & $q_{E,1}$ & $Q_{E,2}$ & $q_{E,2}$ & $Q_{E,3}$& $q_{E,3}$ \\ \hline
    0.1800 & 0.1791 & 0.5000 & 0.4964 & 0.6600 & 0.6574 \\ \hline
    0.3800 & 0.3823 & 0.6200 & 0.6173& 0.7400 & 0.7398 \\ \hline \hline
\end{tabular}
\begin{tabular}{|c|c|c|c|c|c|}
\hline
\multicolumn{6}{|c|} {Simulation of other type of error} \\ \hline \hline
    $P_{1,1}$ & $p_{1,1}$ & $P_{1,2}$ & $p_{1,2}$ & $P_{1,3}$ & $p_{1,3}$  \\ \hline
    0.0217 & 0.0214 & 0.0581 & 0.0570 & 0.0753 & 0.0738 \\ \hline
    0.0476 & 0.0469 & 0.1219 & 0.1203 & 0.1549 & 0.1546 \\ \hline
    0.0789 & 0.0789 & 0.1923 & 0.1918 & 0.2391 & 0.2387 \\ \hline
    0.1176 & 0.1176 & 0.2703 & 0.2690 & 0.3283 & 0.3275 \\ \hline
    0.1667 & 0.1677 & 0.3571 & 0.3546 & 0.4231 & 0.4235 \\ \hline
    0.0501 & 0.0514 & 0.0793 & 0.0808 & 0.0932 & 0.0964  \\ \hline
    0.1061 & 0.1039 & 0.1623 & 0.1585 & 0.1878 & 0.1833 \\ \hline
    0.1691 & 0.1677 & 0.2493 & 0.2499 & 0.2839 & 0.2831 \\ \hline
    0.2405 & 0.2402 & 0.3407 & 0.3366 & 0.3814 & 0.3782 \\ \hline
    0.3220 & 0.3234 & 0.4366 & 0.4380 & 0.4805 & 0.4813  \\ \hline \hline
\end{tabular}
\begin{tabular}{|c|c|c|c|c|c|}
\hline
\multicolumn{6}{|c|}{Simulation of another type of error} \\ \hline \hline
    $P_{2,1}$ & $p_{2,1}$ & $P_{2,2}$ & $p_{2,2}$ & $P_{2,3}$ & $p_{2,3}$  \\ \hline
    0.5233 & 0.5262 & 0.6429 & 0.6433 & 0.7286 & 0.7258 \\ \hline
    0.3279 & 0.3298 & 0.4444 & 0.4492 & 0.5405 & 0.5337 \\ \hline
    0.2215 & 0.2195 & 0.3182 & 0.3200 & 0.4070 & 0.4083 \\ \hline
    0.1546 & 0.1561 & 0.2308 & 0.2370 & 0.3061 & 0.3080 \\ \hline
    0.1087 & 0.1106 & 0.1667 & 0.1684 & 0.2273 & 0.2225 \\ \hline
    0.7438 & 0.7399 & 0.8257 & 0.8196 & 0.8738 & 0.8739  \\ \hline
    0.5634 & 0.5566 & 0.6780 & 0.6731 & 0.7547 & 0.7512  \\ \hline
    0.4294 & 0.4299 & 0.5512 & 0.5550 & 0.6422 & 0.6393  \\ \hline
    0.3261 & 0.3230 & 0.4412 & 0.4382 & 0.5357 & 0.5349  \\ \hline
    0.2439 & 0.2490 & 0.3448 & 0.3420 & 0.4348 & 0.4385  \\ \hline
\end{tabular}
\end{center}
}

\label{simtable 4}
\end{table}

\begin{table}[htb]
\caption{Simulation of proportion of number of correct detections by Bayes and MP tests}
{
\begin{center}
\begin{tabular}{|c|c|c|c|c|c|c|c|}
\hline
\multicolumn{4}{|c|}{$p_1 = 0.9,p_2 = 0.5,p_3 = 0.3, p_w = 0.1, p_c = 0.9$} \\ \hline \hline
    $p_e$ & $l$ & $1-$ type I error (simulated) & Power (simulated) \\ \hline
    0.1 & 5 & 0.9016  & 0.9436 \\ \hline
    0.3 & 5 & 0.9417  & 0.9131 \\ \hline
    0.5 & 5 & 0.9819  & 0.7554 \\ \hline
    0.1 & 20 & 0.9465 & 0.9090 \\ \hline
    0.3 & 20 & 0.9866 & 0.7358 \\ \hline
    0.5 & 20 & 0.9931 & 0.7291 \\ \hline \hline
\multicolumn{4}{|c|}{$p_1 = 0.7,p_2 = 0.3,p_3 = 0.1, p_w = 0.2, p_c = 0.8$} \\ \hline \hline
    0.1 & 5 & 0.5927 & 0.8537   \\ \hline
    0.3 & 5 & 0.8520 & 0.5963   \\ \hline
    0.5 & 5 & 0.9610 & 0.2682   \\ \hline
    0.1 & 20 & 0.8367 & 0.6001   \\ \hline
    0.3 & 20 & 1.0000 & 0.0    \\ \hline
    0.5 & 20 & 1.0000 & 0.0    \\ \hline \hline

\multicolumn{4}{|c|}{$p_1 = 0.9,p_2 = 0.5,p_3 = 0.3, p_w = 0.1, p_c = 0.9$} \\ \hline \hline
    MP test & $\alpha$ & $1-$ type I error (simulated) & Power (simulated)  \\ \hline
     &  0.900   & 0.9014 & 0.9437 \\ \hline
     &  0.950   & 0.9500 & 0.8814 \\ \hline
     &  0.975   & 0.9767 & 0.7897 \\ \hline
     &  0.990   & 0.9906 & 0.6231 \\ \hline \hline
\multicolumn{4}{|c|}{$p_1 = 0.7,p_2 = 0.3,p_3 = 0.1, p_w = 0.2, p_c = 0.8$} \\ \hline \hline
     &  0.900   & 0.8875 & 0.4985 \\ \hline
     &  0.950   & 0.9458 & 0.3296 \\ \hline
     &  0.975   & 0.9735 & 0.1856 \\ \hline
     &  0.990   & 0.9891 & 0.0882 \\ \hline
\end{tabular}
\end{center}
}

\label{simtable 5}
\end{table}

{\it Observations:} A few immediate observations from the theoretical results (which can be verified in Table~\ref{simtable 4}) are as follows:

\begin{enumerate}
\item The simulated and theoretical values of the different errors are close enough; they differ by at most $2\%$.
\item The simulated and theoretical values of type I error of the MP test are approximately same; they differ by at most $0.3\%$.
\item We simulated type II error of the MP test and both the errors for the Bayes test, which is very hard to calculate theoretically.
\item Powers of the Bayes and MP test are very close for the same type I error, e.g. for $p_1 = 0.9,p_2 = 0.5,p_3 = 0.3, p_w = 0.1, p_c = 0.9$ type I error for the MP test and the Bayes test are $0.9014$ and $0.9016$, respectively, and the corresponding type II errors are $0.9437$ and $0.9436$. This indicates that both test are good and equally powerful. In case of the MP test, the type I error (i.e., $1-\alpha$) has to be chosen before the test, but in case of Bayes test if the ratio of the losses is known then type I error is automatically fixed and the overall loss is minimized.
\end{enumerate}

\section{Estimation of the Parameters}
In practice, the problem is that the parameters $p_1, p_2, p_3, p_w$ and $p_c$ may be unknown. We can, however, estimate these parameters through experimentation.

Note that $P(y_{1j}=1| \Event) = p_1$. Hence, $p_1$ is the expected value of $y_{1j}$ given an event. So, we perform the experiment as follows: an event occurs in some node $N$ of the ROI and we count how many $y_{1j}$'s gives value $1$. The proportion of $y_{1j}$'s having value $1$ gives an estimate of $p_1$. We repeat this experiment several times so that the average of the proportions over repeated experiments can be taken as an estimate of $p_1$.

Note that under normal situation, $x_{ij}$ follows $\Ber(p_w)$ for all $i,j$. Hence, $p_w$ is the expected value of $x_{ij}$ given normal situation. So, we perform the experiment by keeping the ROI normal and find the values of $x_{ij}$'s.  The proportion of $x_{ij}$'s having value $1$ gives an estimate of $p_w$. We repeat this experiment several times so that the average of the proportions over repeated experiments can be taken as an estimate of $p_w$.

Similar experiments will give the expected value of $p_2, p_3,$  and $p_c$.

\section{Some Special Cases}

 With fewer sensors, e.g., only center node or center node and four distance-one sensors can detect the event square or when sensors always send the message correctly with probability $1$ or when the sensors can detect the event square without any error, we can simplify and say more about the error probabilities and the tests. In following subsections we discuss some special cases of that nature.

\subsection{When sensors always send the message correctly}
We have, $p_w =0$ and $p_c =1 $. Hence, $d=1$ since $d = p_c - p_w$.

The type I and type II errors for $N_{ij}$ are $1-p_i$ and $0$, respectively, and
$$P_{1,i} = \frac{p_e(1-p_i)}{1-p_ep_i} \approx p_e(1-p_i)(1+p_ep_i)\approx p_e(1-p_i), $$
for small $p_e$ and $P_{2,i} = 0$ for all possible values of $i$, $j$.
$$\mbox{If \ } \left(\frac{p_n}{p_e l}\right)\left(\frac{1}{1-p_1}\right)\left(\frac{1}{1-p_2}\right)^4\left(\frac{1}{1-p_3}\right)^4 > 1 ,$$
then the Bayes test is to reject $H_0$ only when $x_1 = x_2 = x_3 = 0 $.
Otherwise, the Bayes test is to accept $H_0$ for all values of $x_1, x_2, x_3$.
$$\mbox{Therefore, if\ } l > \frac{p_n}{p_e}(1-p_1)^{-1}(1-p_2)^{-4}(1-p_3)^{-4}, $$
the Bayes the test is to reject $H_0$ when $x_1 = x_2 = x_3 = 0 $
Otherwise, the Bayes the test is to accept $H_0$ for all values of $x_1, x_2, x_3$.
$$\mbox{If  \ } (1-p_1)(1-p_2)^4(1-p_3)^4 \leq \a ,$$
then the MP test of size $\a$ is to reject $H_0$ when $x_1 = x_2 = x_3 = 0$. Otherwise, the MP test of size $\a$ is to accept $H_0$ for all values of $x_1, x_2, x_3$.

\subsection{When the sensors can detect the event square without any error}

We have the detection probability is $1$, i.e., $p_i=1$ for $i=1,2,3.$ Then the type I and type II error for $N_{ij}$ are $1-p_c$ and $p_w$, respectively.
$$P_{1,i} = \frac{p_e(1-p_w-d)}{1-p_w-p_ed} \mbox{\ and\ } P_{2,i} = \frac{(1-p_e)p_w}{p_w+p_ed}$$
for all possible values of $i$ and $ j.$

The Bayes test is to reject $H_0$ when
$$x_1 + x_2 + x_3 < \frac{ln\left(\frac{1-p_e}{lp_e} \right) + 9ln\left(1 + \frac{d}{1-p_c}\right)}{ln\left(1+ \frac{d}{p_w(1-p_c)}\right)}.$$

The MP test of size $\a$ is to reject $H_0$ when $x_1 + x_2 + x_3 < \l$
and reject $H_0$ with probability $k$, when $x_1 + x_2 + x_3 = \l$.

$\l$ and k can be found from the relation
$$\Pr(x_1 + x_2 + x_3 < \l ) + k\Pr(x_1 + x_2 + x_3 = \l ) = \a, $$

where $x_1 + x_2 + x_3$ follows $\Bin (9,p_c).$

\subsection{When one center and four distance-one sensors can detect the event square, i.e., $p_3=0$}

If sensors have less sensing power (i.e., small sensing radius), then small numbers of sensors can detect the event square. Assume that only center and four adjacent sensors can detect the event square.
Then, we consider only five squares: one center square and four one-distanced adjacent squares.

The Bayes test is to reject $H_0$ when $\l_1x_1 + \l_2x_2 < 1$\\
where,  for $i=1,2$, $\l_i = $
$$ \frac{ln\left(1+ \frac{p_id}{p_w(1-p_w-p_id)}\right)}{ln(\frac{1-p_e}{lp_e}) + ln\left(1 + \frac{p_1d}{1-p_w - p_1d}\right) + 4ln\left(1 + \frac{p_2d}{1-p_w - p_2d}\right)}.$$

The MP test of size $\a$ is to reject $H_0$ when $t_1x_1 + t_2x_2 < \l$ \\
and reject $H_0$ with probability $k$, when $t_1x_1 + t_2x_2 = \l$,

$$\mbox{where,\ } t_i = ln\left(1+ \frac{p_id}{p_w(1-p_w-p_id)}\right), i=1,2.$$

$\l$ and $k$ can be find from the relation
$$\Pr(t_1x_1 + t_2x_2 < \l ) + k\Pr(t_1x_1 + t_2x_2 = \l ) = \a, $$

where $x_1$ follows $\Ber (p_w + p_1d)$, $x_2$ follows $\Bin (4,p_w + p_2d)$ and they are independent when $H_0$ is true.

\section{When Sensors are Placed at the Centers of Regular Hexagons}
  
In this section, we assume ROI is partitioned into congruent regular hexagons (which are known as cells) with side $a$, i.e., we can think ROI as a hexagonal grid with regular hexagonal cells. We consider that sensors are placed previously at the center of each cell of the hexagonal grid. We assume that the sensor network covers the entire ROI. Instead of three detection probability (as in the case of square grid), we assume there are two detection probability $p_1, p_2$, where $p_1 > p_2$. Note that there are six adjacent nodes of a particular node.

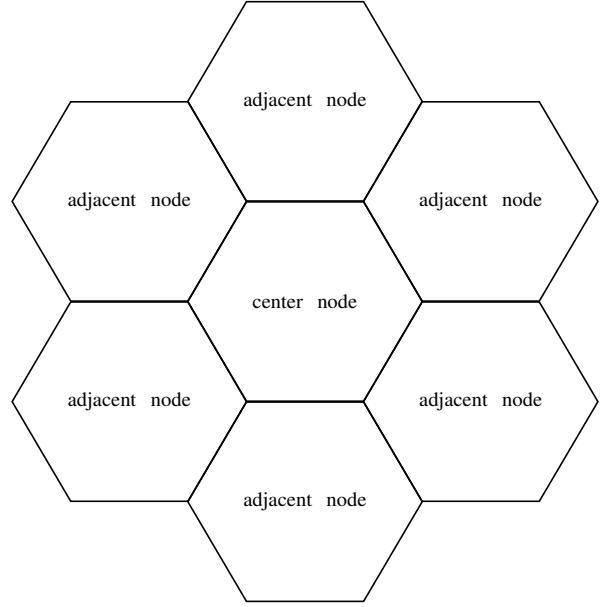
\begin{figure}
\resizebox{8 cm}{8 cm}{

\begin{tikzpicture}[scale=2]

\node [draw, thick, minimum size=4cm, regular polygon, regular polygon sides=6] at (2,2) {center \ node};
\node [draw, thick, minimum size=4cm, regular polygon, regular polygon sides=6] at (2,3.74) {adjacent \ node};
\node [draw, thick, minimum size=4cm, regular polygon, regular polygon sides=6] at (2,0.26) {adjacent \ node};
\node [draw, thick, minimum size=4cm, regular polygon, regular polygon sides=6] at (0.5,2.87) {adjacent \ node};
\node [draw, thick, minimum size=4cm, regular polygon, regular polygon sides=6] at (0.5,1.13) {adjacent \ node};
\node [draw, thick, minimum size=4cm, regular polygon, regular polygon sides=6] at (3.5,2.87) {adjacent \ node};
\node [draw, thick, minimum size=4cm, regular polygon, regular polygon sides=6] at (3.5,1.13) {adjacent \ node};
\end{tikzpicture}
}
\caption{Nodes placed  in hexagonal grid when ROI partitioned in to regular hexagons}
\end{figure}

We define $N_{ij}, x_{ij}, y_{ij}, x_i, p_e, p_n, p_c, p_w, t_i$ as in the case of square grid.

Hypotheses are also same as in the case of square grid.

The Bayes test is to reject $H_0$ when
$$\l_1x_1 + \l_2x_2 < 1,$$
$$\mbox{where,}\ \l_i = \frac{ln\left(1+ \frac{p_id}{p_w(1-p_w-p_id)}\right)}{ln(c)}, \mbox{\ for} \ i=1,2; $$
with $c = $
$$\left(\frac{1-p_e}{lp_e}\right)\left(1 + \frac{ p_1d}{1-p_w - p_1d}\right)\left(1 + \frac {p_2d}{1-p_w - p_2d}\right)^6.$$

The MP test of size $\a$ is to reject $H_0$ when
$$t_1x_1 + t_2x_2 < \l$$
and reject $H_0$ with probability $k$, when
$$t_1x_1 + t_2x_2 = \l,$$
$$\mbox{where,}\ t_i = ln\left(1+ \frac{p_id}{p_w(1-p_w-p_id)}\right), i=1,2.$$
$\l$ and $k$ can be found from the relation
$$\Pr(t_1x_1 + t_2x_2 < \l ) + k\Pr(t_1x_1 + t_2x_2 = \l ) = \a, $$
$x_1$ follows Ber ($p_w + p_1d)$, $x_2$ follows Bin ($6,p_w + p_2d)$ and $x_1$ and $x_2$ are independent under $H_0$.

\section{Concluding Remarks and Future Work}

In this paper, we have considered the problems for fault detection in wireless sensor network (WSN). We partition the ROI as a rectangular grid with square cells. We discuss how to address both the noise-related measurement error and sensor fault simultaneously in fault detection where the sensors are placed at the centers of square cells of the ROI and the event occurs at only one square of the grid. We have also considered the ROI as partitioned into regular hexagonal cells and do the same analysis. We have proposed fault detection schemes that explicitly introduce the error probability into the optimal event detection process. We developed the schemes under 1) classical hypothesis testing and 2) Bayes test. We have identified and analyzed all the situations in which these tests are effective and cases where they are not applicable.

We observed that type I and type II errors decrease when $p_i$ and $ p_c$ increase and increase when $ p_w$ increases. When $p_e$ increases, type I errors increase but type II errors decrease. If detection probability $p_i$ is low then type I error is close to $p_e$. If $p_e$ is close to $0.5$ then type I error is close to $p_e$ which means that there is no use of sensors; in that case, we have to use sensors with high detection probability.

In future, we plan to do the following:
\begin{enumerate}
\item Develop schemes to find which particular square is the event square.
\item Develop schemes to identify the event square when there are more than one event squares.
\item Develop schemes to find and isolate the dead and the faulty sensors, i.e., the sensors which are sending the false information to the base station.
\item We may assume that sensors can detect different type of events; thus, response of sensors may not be simply binary.
\item We may assume that sensors can measure distance, direction, speed, humidity, wind speed, soil makeup, temperature, etc. and send the measurement of continuous type variables.
\end{enumerate}

{\bf Acknowledgments}
The authors would like to thank Prof. Anup Dewanji of Indian Statistical Institute, Kolkata, for their valuable suggestions towards the technical contribution of this paper. Mrinal Nandi would also like to thank the Department of Information Technology, Govt. of India, for partially supporting this project.

\end{document}